\documentclass[aps,prb,twocolumn,superscriptaddress,showpacs]{revtex4}

\usepackage[dvips]{graphicx}

\begin{document}

\title{Phonon-Induced Quantum Magnetic Deflagration in Mn$_{12}$}


\author{A. Hern\'andez-M\'inguez}
\author{F. Maci\`a}
\author{J. M. Hernandez}
\author{J. Tejada}
\affiliation{Departament de F\'isica Fonamental, Facultat de
F\'isica, Universitat de Barcelona\\Avda. Diagonal 647, Planta 4,
Edifici nou, 08028 Barcelona, Spain}

\author{P. V. Santos}
\affiliation{Paul-Drude-Institut f\"ur Festk\"orperelektronik,
Hausvogteiplatz, 5-7, 10117 Berlin}


\date{\today}

\begin{abstract}
A comprehensive set of experiments on the effect of high-frequency
surface acoustic waves, SAWs, in the spin relaxation in
Mn$_{12}$-acetate is presented. We have  studied  the quantum
magnetic deflagration induced by SAWs under various experimental
conditions extending the data shown in a very recent paper [A.
Hern\'andez-M\'inguez \textit{et. al.}, Phys. Rev. Lett.
\textbf{95}, 217205 (2005)]. We have focused our study on the
dependence of both the ignition time and the propagation speed of
the magnetic avalanches on the frequency, amplitude, and duration
of the SAW pulses in experiments performed under different
temperatures and external magnetic fields.
\end{abstract}


\maketitle

\section{Introduction}
Molecular nanomagnets are well-defined discrete molecules
consisting of several transition metal ions interacting through
organic and/or inorganic ligands. The so called single molecule
magnets, SMMs, are characterized by two energy scales: i) the
strong exchange interactions between metal ions within a molecule,
and ii) an anisotropic spin-orbit coupling that results in
pronounced magnetic anisotropy. Some of these SMMs have large spin
($S$) values as well as a high energy anisotropy barrier to
prevent spontaneous magnetization reversal at low
temperatures.\cite{SESSOLI} These well characterized objects have
an intermediate size between microscale and macroscale and thus
allow for the study of phenomena at the border between macroscopic
quantum tunnelling and the conventional quantum mechanics of spin.
The first unambiguous evidence for resonant magnetization
tunnelling in molecular magnets came from the stepwise magnetic
hysteresis experiments at different temperatures performed on a
Mn$_{12}$-acetate sample, which has $S=10$ and an anisotropy
barrier height \mbox{$U=DS_{z}^{2}=60$ K}.\cite{TEJADA} Since
then, numerous experiments have been performed  on several SMMs
demonstrating the existence  of very interesting phenomena,  which
are well accounted for by theoretical calculations. For a review
see Refs.~\onlinecite{ rev:Gatteschi, rev:Friedman, rev:delBarco}.

To  understand the physics of the SMMs, let us assume the simplest
Hamiltonian  with the magnetic field applied parallel to the easy
anisotropy axis:
\begin{equation}
\mathcal{H} = -DS_{z}^{2} - H_{z}S_{z}
\end{equation}

\noindent This Hamiltonian yields pairs of degenerate levels for
fields $H_z=nD$ with $n=0,1,\ldots, S$. When all molecules occupy
the spin states with a negative magnetic quantum number $m$, the
sample is said to be magnetized in the negative direction. By
applying a magnetic field in the positive $z$ direction, the
molecules eventually relax to the spin states with positive $m$.
This process may occur via thermal transitions over the energy
barrier or through quantum tunnelling between the states on
different sides of the energy barrier. The tunnelling adds to the
thermal activation when the levels at the two sides of the barrier
are on resonance, that is at $H_{z}=nD$. To account for
tunnelling, one must include additional terms in the spin
Hamiltonian, which do not commute with $S_z$. These terms may be
associated with transverse anisotropy, like quadratic and
higher-order terms in $S_x$ and $S_y$ or the transversal Zeeman
energy term  $-H_xS_x$.

From the experimental point of view, there are two approaches to
study spin tunnelling: by  1) measuring  the magnetic relaxation
at fixed temperature and magnetic field or by 2) sweeping the
magnetic field through the resonant field  and detecting the
fraction of molecules, $\eta$, that change their magnetic moment.
$\eta$ depends on the overlap of the wave functions of the
resonant spin states at the two sides of the barrier, which
defines the tunnelling splitting, $\Delta$, as well as on the
sweep rate of the magnetic field energy, $v$,  according to the
expression

\begin{equation}
\eta=1-\exp(-\frac{\pi\Delta^{2}}{2\hbar v})
\end{equation}

Since the 1990s,~\cite{art:Fominaya, art:delBarco} it has been
known that for fast sweeping rates, the magnetic reversal in large
crystals takes place in a short time scale (typically less than
\mbox{1 ms}) through so-called magnetic avalanches. In this
process, the initial relaxation of the magnetization towards the
direction of the magnetic field results in the release of heat
that further accelerates the magnetic relaxation. Recent local
magnetic measurements on Mn$_{12}$-acetate crystals have
demonstrated that during an avalanche the magnetization reversal
occurs inside a narrow interface that propagates through the
crystal at a constant speed of a few meters per second.\cite{YOKO}
This phenomenon has been named ``magnetic deflagration'' because
of its parallelism to classical chemical deflagration.
Deflagration is a subsonic combustion that propagates through
thermal conductivity (a steady flame heats the next layer of cold
material and ignites it). Classically, combustion describes the
exothermic chemical reaction between a substance and a gas
(usually O$_2$). The gas oxidize the substance (the fuel), leading
to heat release. During magnetic deflagration, the magnetic field
``oxidizes''  spins originally in a meta-stable state and makes
them flip. The released heat then feedbacks the magnetization
reversal process.

Recently, a novel method for the controlled ignition of avalanches
at constant magnetic field by means of surface acoustic waves,
SAWs, has been  reported.\cite{AHM} These acoustomagnetic
experiments are performed by using hybrid piezoelectric
interdigital transducers deposited on a LiNbO$_3$ substrate, with
the Mn$_{12}$ single crystals directly glued onto the
piezoelectric.\cite{AHM,JMH} The investigations have clearly shown
that the propagation speed of the avalanches exhibits maxima for
magnetic fields corresponding to the tunnelling resonances of
Mn$_{12}$-acetate. The results suggest, therefore, a novel
physical phenomenon: deflagration assisted by quantum tunnelling.
In this paper, we study these well-controlled ignited magnetic
avalanches assisted by spin tunnelling in Mn$_{12}$-acetate under
various experimental conditions.

\section{Experimental set up}

\begin{figure}
\begin{center}
\includegraphics[width=\columnwidth]{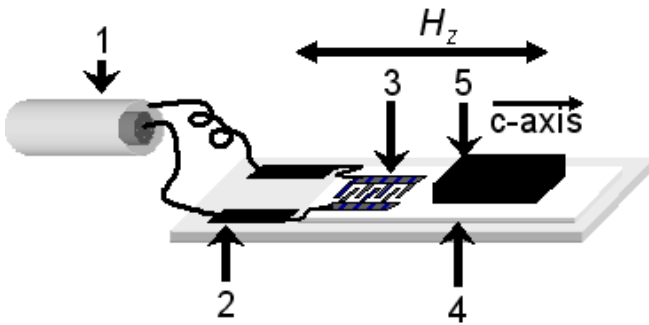}
\caption{Experimental setup. 1-coaxial cable; 2-conducting
stripes; 3-interdigital transducers (IDT); 4-LiNbO$_3$ substrate;
5-Mn$_{12}$ crystal. The $c$-axis of the Mn$_{12}$ crystal is
oriented parallel to the applied magnetic field $H_z$.
\label{fig:setup}}
\end{center}
\end{figure}

The acoustomagnetic experiments were performed by using hybrid
piezoelectric interdigital transducers (IDT) deposited on  128
YX-cut  LiNbO$_{3}$ substrates\cite{PIEZO} with dimensions
$4\times12\times1$ mm$^{3}$. Several single crystals of
Mn$_{12}$-acetate with average dimensions of
\mbox{$2\times.5\times.5$ mm$^{3}$} were studied. Each crystal was
measured independently. The sample was glued directly onto the IDT
using a commercial silicon grease. The microwaves for the SAW
generation were transported to the transducers by coaxial cables
(cf. Fig.~\ref{fig:setup}). Experiments have been carried out
using a commercial rf-SQUID Quantum Design magnetometer at
temperatures between 2 and 2.7~K.

\begin{figure}
\begin{center}
\includegraphics[width=\columnwidth]{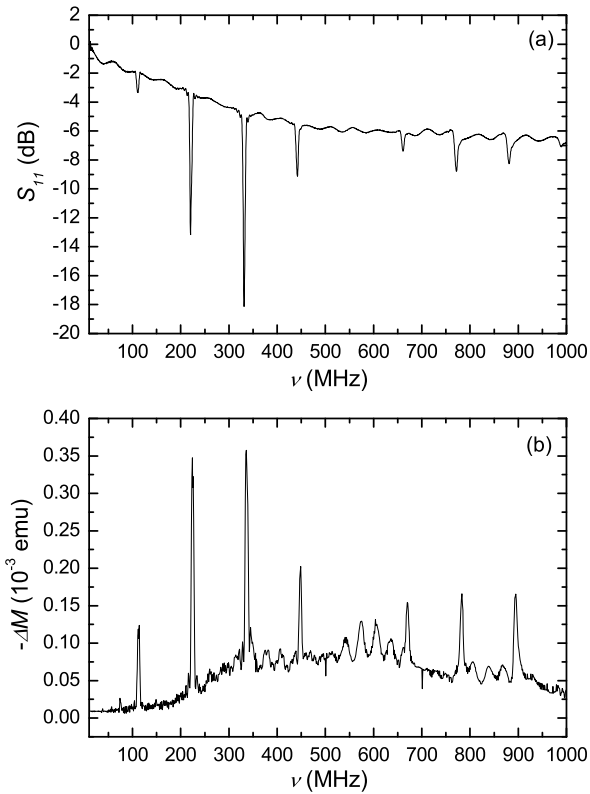}
\caption{(a) Reflection coefficient, $S_{11}$, of  the
LiNbO$_3$/Mn$_{12}$ hybrids (including the effects of the coaxial
assembly used to feed the microwave power) and (b) magnetization
variation measured by the magnetometer as a function of the
frequency of the microwave pulses sent to the interdigital
transducer (IDT). The temperature and applied magnetic field are
$T=6$~K and $H=1$~T, respectively.\label{fig:S11&DM}}
\end{center}
\end{figure}

The frequency response of the IDTs was measured using  an Agilent
network analyzer. The frequency dependence of the reflection
coefficient $S_{11}$, displayed in Fig.~\ref{fig:S11&DM}(a), shows
that the coaxial cables introduce an attenuation smaller than
10~dB. The IDT generates SAWs at multiple harmonics of the
fundamental frequency of \mbox{111 MHz} up to a maximum frequency
of approximately \mbox{1.5 GHz}. Next to the $S_{11}$
determination, we also measured the frequency dependence of the
magnetization changes induced by the SAWs in the superparamagnetic
regime of Mn$_{12}$ (i.e., for temperatures $>3$~K). In this case,
the microwave pulses used to excite the IDT were generated by a
commercial Agilent signal generator,  which allows the selection
of the shape, duration, and energy of the pulses in the frequency
range between \mbox{250 kHz} to \mbox{4 GHz}. Most of the
experiments were performed using rectangular microwave pulses of
duration \mbox{10~$\mu$s~-~10~ms}. Fast magnetization measurements
(time resolution of \mbox{1 $\mu$s}) were carried out at constant
temperature and  magnetic field by continuously recording the
voltage variation detected by the rf-SQUID. The negative
magnetization changes ($-\Delta M$, see Fig.~\ref{fig:S11&DM}b)
show peaks at the resonance frequencies of $S_{11}$, thus clearly
demonstrating that they are induced by the acoustic field. The
magnitude of the magnetization variation for each peak depends on
how much acoustic energy is absorbed by the Mn$_{12}$ single
crystal. During the experiments, the temperature of the IDT
attached to the sample and the temperature of the helium gas that
provided heat exchange were independently monitored.\cite{JMH}

To study magnetic avalanches, we first saturated the sample
magnetization at -2~T (i.e., below the blocking temperature)  and
then swept the magnetic field at a constant rate of 300 Oe/s up to
a predefined value $H$. A few seconds later, we applied a
rectangular microwave pulse with well-controlled frequency,
energy, and duration to excite the SAWs and trigger the magnetic
avalanche. These experiments have established a new method for
igniting magnetization avalanches in molecular magnets with total
control of the magnetic field and initial temperature.
Furthermore, by varying the frequency, duration, and nominal power
of the pulse we have been able to study the dynamics of the
avalanche ignition process.

\section{Discussion}
\subsection{Time evolution of the avalanches}

Figure~\ref{fig:tev} shows the time evolution of the magnetization
during avalanches ignited at $T=2$~K with SAWs pulses of
$t_{p}=5$~ms at different magnetic fields. As the whole experiment
is performed in a time on the order of a few seconds (i.e., much
shorter than the relaxation time in the absence of acoustic
excitation), it can be carried out at any temperature below the
blocking temperature $T_B=3$~K. From the magnetic point of view,
the important experimental fact is that since the magnetization is
measured with a resolution time of 1~$\mu$s, we have been able to
detect three different stages during magnetization reversal:

\begin{figure}
\begin{center}
\includegraphics[width=\columnwidth]{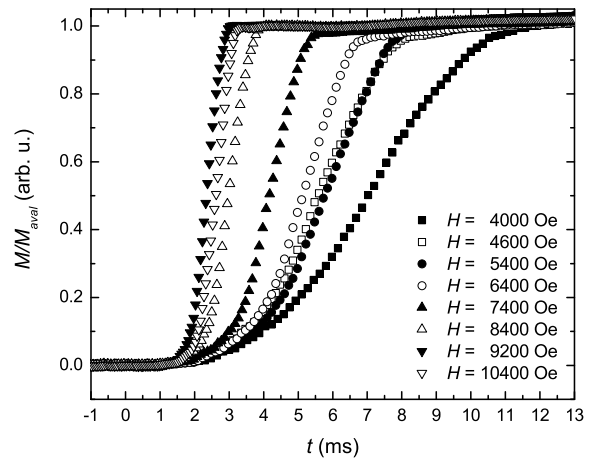}
\caption{Time evolution of the magnetization, $M$,  for different
magnetic fields at $T=2$~K. The magnetization is normalized to the
total magnetization change during the avalanche,
$M_\mathrm{aval}$. The duration of the SAWs pulse is $t_p=5$~ms
and the frequency is $\nu=449$~MHz. $t=0$~ms corresponds to the
instant at which the microwave pulse is applied. The results show
that the rate of change of the magnetization peaks at the resonant
fields of  $H=4600$~Oe and $H=9200$~Oe.\label{fig:tev}}
\end{center}
\end{figure}

\begin{enumerate}

\item The first stage corresponds to the time elapsed between the
application of the microwave pulse and the first detection of
magnetization changes. During this time, the SAWs thermalize and
the magnetization reversal process nucleates within a small region
of the Mn$_{12}$ crystal. To define this nucleation time, we have
adopted the criterium that it corresponds to the time interval
between the application of the SAWs pulse and the detection of the
first magnetization change.

\item The second stage corresponds to the time interval between
nucleation and the formation of a stable deflagration front
propagating with constant velocity through the crystal.\cite{YOKO}
During this time, the nucleation bubble containing reversed spins
becomes larger and larger until a steady interface is formed and
it propagates along the sample (the ``flame'' in magnetic
deflagration). The heat energy liberated during spin reversal
corresponds to the Zeeman energy $\Delta E=g\mu_BH_z\Delta S$ per
spin. The temperature increase induced by the spin reversal leads
to the expansion of the nucleation bubble across the width of the
Mn$_{12}$ crystal (cf. Fig.~\ref{fig:setup}). In contrast to the
subsequent steady-state propagation along the $c$-axis, this
propagation of the front flame depends on the geometry of the
crystal and on its magnetic history. For instance, the time to
reach that flame will depend on the width of the sample. We call
the {\it ignition time}, $t_\mathrm{ign}$, the total duration of
these two first steps.

\item The third stage corresponds to the so-called magnetic
deflagration, where the flame propagates at constant velocity
along the $c$-axis of the Mn$_{12}$ single crystal. During this
time, the rate of magnetization variation is constant and the
flame front has a finite width $\delta$. Contrary to the previous
stage, the temperature remains constant at value $T_f$
considerably greater than those achieved during the flame
formation.
\end{enumerate}

\begin{figure}
\begin{center}
\includegraphics[width=\columnwidth]{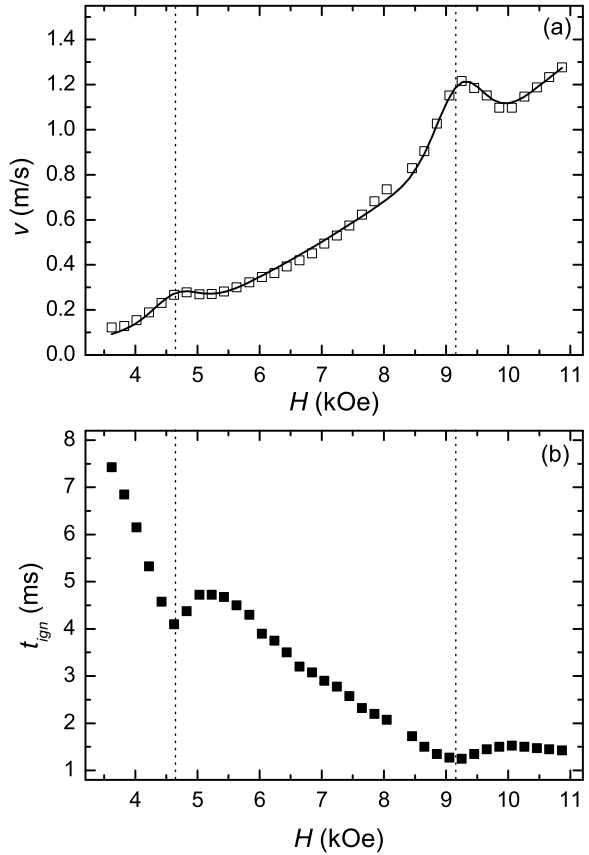}
\caption{(a) Velocity and (b) ignition time ($t_\mathrm{ign}$) of
the deflagration front as a function of  magnetic field recorded
at $T=2$~K by using $t_p=5$~ms SAWs pulses with a frequency
$\nu=449$~MHz. The vertical dotted lines mark the resonant field
values. The solid line displays a fit to
Eq.~\ref{Eq3}.\label{fig:vel&tign}}
\end{center}
\end{figure}

Figure~\ref{fig:vel&tign}(a) shows the velocity $v=l/\Delta t$ of
the deflagration front deduced from the magnetization data
displayed in Fig.~\ref{fig:tev}. Here, $l$ is the length of the
crystal and the avalanche time, $\Delta t$, is defined  as the
time needed by the magnetization to change between the 20\% and
the 80\% of the total variation, $M_{aval}$. The velocity $v$
increases with the applied magnetic field and presents peaks at
the resonant field values. The dependence on magnetic field is
well-fitted (see solid line in Fig.~\ref{fig:vel&tign}(a)) by the
law:~\cite{YOKO}

\begin{equation}
v=\sqrt{\frac{\kappa}{\tau_0}}\exp{\left[-\frac{U(H)}{2k_BT_f}\right]}
\label{Eq3}
\end{equation}

\noindent where $\kappa$ is the thermal diffusivity. $U(H)$,
$\tau_0$ and $T_f$ are, respectively, the energy barrier, the
attempt frequency, and the temperature of the ``flame'', which are
related to the ``chemical reaction time'' $\tau$  by the
expression $\tau=\tau_0\exp\left[U(H)/k_BT_f\right]$. In the case
of Mn$_{12}$, $\kappa\sim10^{-5}$~m$^2/s$, $\tau_0\sim10^{-7}$~s,
and the field dependence of the energy barrier, $U(H)$, is well
known.\cite{art:delBarco, art:Sales} The temperature of the flame
increases linearly with  magnetic field. According to the fitting,
the temperature of the flame front is about 6.8~K for $H=$4600~Oe,
and increases to 10.9~K for $H=$9200~Oe.

Figure.~\ref{fig:vel&tign}b shows the dependence of the ignition
time (i.e., the time delay between the application of the
microwave pulse and the observation of a steady  flame front) on
magnetic field. The ignition time decreases with  increasing
magnetic field, and shows minima at the resonant fields. The later
are a consequence of the enhanced spin reversal probability at the
resonant fields. In other words, the ignition time reproduces the
dependence of the effective barrier height for magnetic
transitions on the magnetic field.

\subsection{The applied acoustic energy}
\label{subsec:energy}

To analyze the dependence of the avalanches on the acoustic energy
supplied by the SAWs, we applied SAWs pulses of different
durations, $t_{p}$, and power, $P$, and the time evolution of the
magnetization for magnetic fields at and out of resonance was
recorded.

\begin{figure}
\begin{center}
\includegraphics[width=\columnwidth]{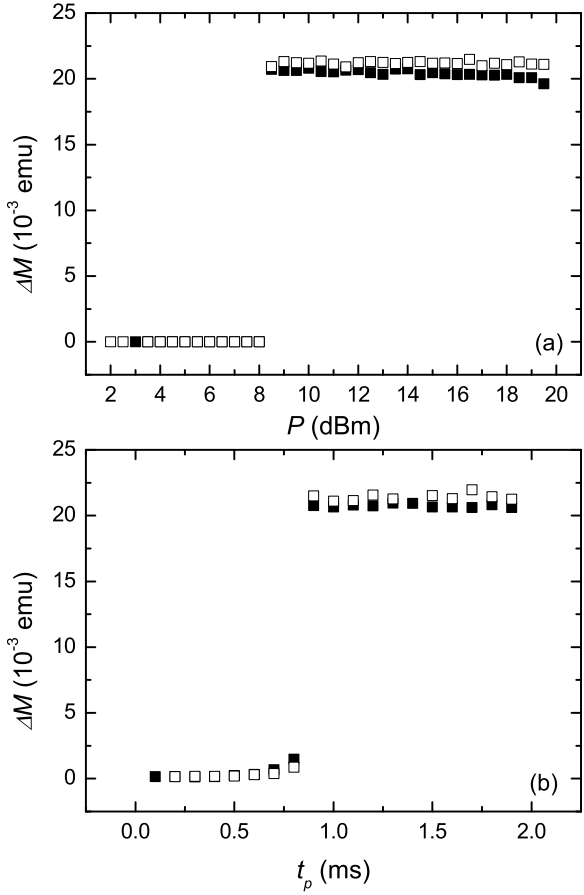}
\caption{Dependence of magnetization changes during the avalanche
on the (a) power ($P$, recorded for $t_{p}=1$~ms) and (b) duration
($t_p$, recorded for for $P=20$ dBm) of the SAWs pulses. The
measurements were carried out at $T=2$ K under magnetic fields of
$H=8800$ Oe (solid squares) and $H=11000$ Oe (open
squares).\label{fig:dMvstp&P}}
\end{center}
\end{figure}

A very interesting point is that, at a given field and
temperature, there is a minimum value for the acoustic energy $P
t_p$ required to trigger the avalanche. As illustrated in
Fig.~\ref{fig:dMvstp&P}(a) and Fig.~\ref{fig:dMvstp&P}(b), this
threshold energy can be surpassed by varying either the amplitude
or the duration of the acoustic pulses. The acoustic energy
supplied by the piezoelectric in excess of the threshold energy
for ignition has no further influence on the avalanche dynamics,
which becomes determined only by temperature and by the amplitude
of the magnetic field.

These results can be easily understood if we consider that the
threshold energy is the minimal energy required  to create the
initial bubble of reversed spins. Once the nucleation process is
overcome, the energy released by spin reversal is much greater
than the acoustic energy delivered by the SAWs. This process is
analogous to the ignition of a combustion process: the material to
be burned has to be externally heated only until the combustion
begins. After that, the amount of energy released by the
exothermic reaction is large enough to maintain the combustion
front propagating through the material.

\subsection{Temperature dependence}
\label{subs:Tdep}

\begin{figure}
\begin{center}
\includegraphics[width=\columnwidth]{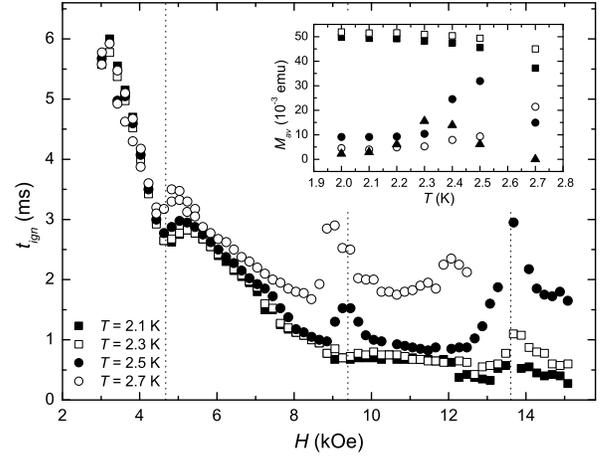}
\caption{Ignition time versus applied magnetic field, $H$, for
different temperatures. The duration of the pulse and frequency
are $t_p=10$~ms, $\nu=224$~MHz. The inset shows the magnitude of
the avalanche for different temperatures and magnetic fields:
$H=4800$~Oe (solid squares), $H=6400$~Oe (open squares),
$H=9200$~Oe (solid circles), $H=11000$~Oe (open circles),
$H=14000$~Oe (solid triangles).\label{fig:tignvsT}}
\end{center}
\end{figure}

\begin{figure}
\includegraphics[width=\columnwidth]{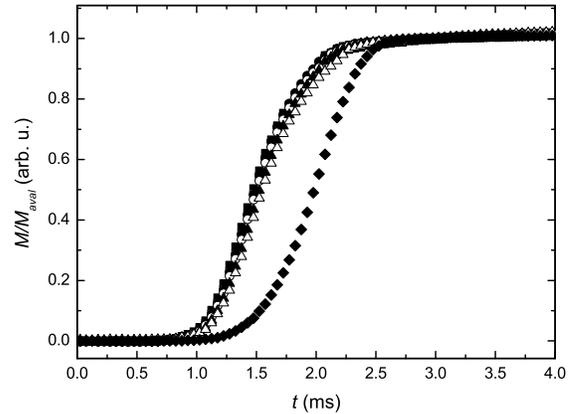}\\
\caption{Time evolution of magnetization for different
temperatures under  an applied field  $H=8000$~Oe. The values of
the initial temperature are: $T=2.0$~K (solid squares), $T=2.1$~K
(open squares), $T=2.2$~K (solid circles), $T=2.3$~K (open
circles), $T=2.4$~K (solid triangles), $T=2.5$~K (open triangles),
$T=2.7$~K (solid rhombus).}\label{fig:tevvsT}
\end{figure}

In order to study the dependence of the deflagration on the
initial amount of reversed spins, we have produced magnetic
avalanches at different temperatures and magnetic fields by
applying acoustic pulses of fixed duration and amplitude.
Figure~\ref{fig:tignvsT} illustrates the typical behavior found at
low temperatures (i.e., below 2.3~K). Under these conditions, the
ignition time is temperature independent and decreases
exponentially with applied magnetic field, with minima at the
resonant field values.

Nevertheless, as the blocking temperature is approached, the
ignition time becomes sensitive not only to the magnetic field but
also to temperature. In particular, the features associated with
the second and third resonances (at $H=9.5$~kOe and $H=13.5$~kOe,
respectively) change from minima in the $t_\mathrm{ign}$ traces
recorded at low temperatures ($T<2.3$~K) to maxima in traces
measured above 2.3~K. At first sight, one should attribute this
behavior to the enhanced relaxation that must occur at the
resonant fields before the SAWs pulse is applied: since the
concentration of metastable spins reduces, it takes longer to
create the nucleation bubble required to launch the front flame.
In fact, the total magnetization change during the avalanche,
$M_{aval}$, decreases slightly with temperature at low magnetic
fields, as indicated in the inset of Fig.~\ref{fig:tignvsT}.
However, at high fields and temperatures above $2.3$~K, $M_{aval}$
presents an anomalous behavior, which has been consistently
observed in several crystals. This behavior is presently not
understood and makes unclear the relationship between $t_{ign}$
and the initial magnetic state of the sample in this temperature
range.

Figure.~\ref{fig:tevvsT} shows the time evolution of the
magnetization for avalanches ignited at different temperatures
under a fixed magnetic field \mbox{$H=8$~kOe}. Like the ignition
time, the velocity of the avalanche depends weakly on the
temperature for low magnetic fields, but slowers down at high
fields and temperatures.

\subsection{Dependence on SAWs frequency}

\begin{figure}
\begin{center}
\includegraphics[width=\columnwidth]{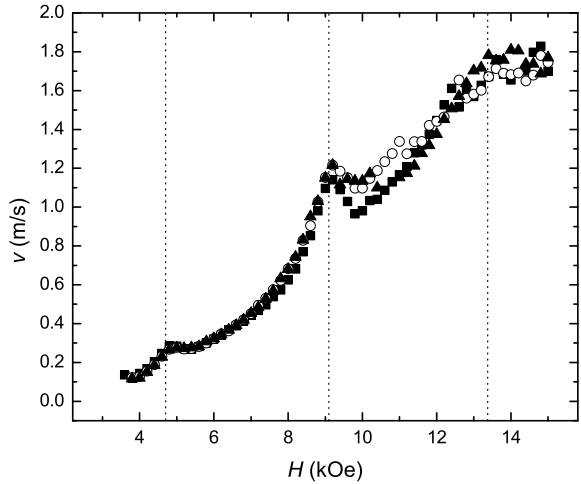}
\caption{Velocity of the  flame front, $v$,  versus applied
magnetic field, $H$, for different SAWs frequencies. The
temperature and SAWs pulse duration are  $T=2$ K and $t_p=5$ ms,
respectively. The SAWs  frequencies and powers are: $\nu=224$ MHz
and $P=13.7$ dBm (solid squares); $\nu=449$ MHz and $P=15.2$ dBm
(open circles); $\nu=895$ MHz and $P=20$ dBm (solid
triangles).\label{fig:velvsfrec}}
\end{center}
\end{figure}

\begin{figure}
\begin{center}
\includegraphics[width=\columnwidth]{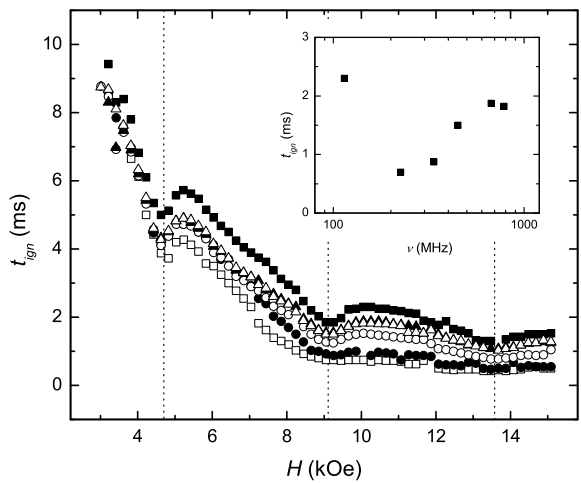}
\caption{Ignition time ($t_\mathrm{ign}$) versus applied magnetic
field, $H$, and different frequencies of the SAWs. The temperature
is $T=2.1$ K and $t_p=10$ ms. The values of frequency are:
$\nu=114$ MHz (solid squares), $\nu=225$ MHz (open squares),
$\nu=336$ MHz (solid circles), $\nu=449$ MHz (open circles),
$\nu=671$ MHz (solid triangles) and $\nu=783$ MHz (open
triangles). The inset shows the ignition time for different
frequencies at $H=10200$~Oe.\label{fig:tignvsfrec}}
\end{center}
\end{figure}

To study the effect of the SAWs frequency on the magnetic
avalanches, it is necessary to ensure that, for each frequency, we
couple the same amount of acoustic energy to the Mn$_{12}$
crystals. For that purpose, we fixed the duration $t_{p}$ of the
SAWs pulses and selected their amplitude in order to produce the
same magnetization variation when the sample is in the
superparamagnetic regime. The results are shown in
Figs.~\ref{fig:velvsfrec} and~\ref{fig:tignvsfrec}. As expected,
the propagation velocity (Fig.~\ref{fig:velvsfrec}) is essentially
independent of  the frequency of the SAWs,  since it is primarily
determined by the internal energy liberated during spin reversal.
In contrast, the ignition time (Fig.~\ref{fig:tignvsfrec}), which
depends on the external energy supplied by the SAWs, shows a
strong frequency dependence. Furthermore, for each  magnetic
field, the ignition time shows a non-monotonic behavior with
frequency, with a minimum at 225~MHz.

\section{Conclusion}

We have investigated the dynamics of magnetic avalanches induced
by acoustic waves in Mn$_{12}$ crystals. The magnetic avalanche
becomes a fully deterministic process under acoustic excitation.
When the acoustic power exceeds the threshold for the nucleation
of the avalanche process, the avalanche dynamics can, therefore,
be investigated for different values of the magnetic field and
temperature. Well below the blocking temperature, the velocity of
the deflagration front of reversing magnetization and the ignition
time do only depend on the applied magnetic field. The ignition
time depends also on the frequency of the SAWs.

\begin{acknowledgments}
We are indebted to  W. Seidel and S. Krauss for the fabrication of
the transducers for SAWs generation. A. H-M. thanks the Spanish
Ministerio de Educación y Ciencia for a research grant. J. M. H.
thanks the Ministerio de Educación y Ciencia and the University of
Barcelona for a Ramón y Cajal research contract. F. M. and J. T.
thank SAMCA Enterprise for financial support.
\end{acknowledgments}

\end{document}